# Slowing down of charged particles in the dusty plasmas with a non-thermal velocity α-distribution

Wang Yu, Du Jiulin

*Department of Physics, School of Science, Tianjin University, 300072，Tianjin, China*

**Abstract** The slowing down of a charged particle beam passing through the dusty plasma with a non-thermal velocity α-distribution is studied. By using the Fokker-Planck collision theory, we derive the deceleration factor and slowing down time and make the numerical analyses. We show that the non-thermal velocity α-distributions of the plasma components have a significant effect on the slowing down. With increase of the mean velocity, the deceleration factor increases rapidly, reaches a peak and then decreases gradually. And the entire peak of the deceleration factor moves generally to the right with the increase of the α-parameter. The slowing down time decreases with the increase of the non-thermal α-parameter, and so the slowing down time in the non-thermal dusty plasma is generally less than that in a Maxwellian one.

**1. Introduction**

In the astrophysical and space plasmas, dusty plasmas are ubiquitous, such as those in interstellar clouds, periplanetary clouds, interplanetary space, comets, planetary rings, the Earth's atmosphere and the lower ionosphere etc. In the dusty plasmas, dust particles are not electrically neutral; they may carry negative or positive charges under the joint interactions with the surrounding plasma environment. Dusty plasma can also be in the laboratory plasma environments. Therefore, the properties about dusty plasmas have been widely studied in recent many years.

The slowing down of charged particles in the dusty plasma means that when a charged particle beam flies into the plasma, it will decelerate due to the resistance of plasma components. Suppose a charged beam of particles impinges upon the background dusty plasma. The slow down occurs due to collisions between this beam and the background plasma as it will loses momentum due to the collisions with the particles in the background plasma. In addition, there should be a broadening of the beam particle velocity distribution since the collisions will also tend to randomize the velocity of the beam particles. Meanwhile, the background plasma should be heated and also gain momentum due to the collisions. Eventually, the beam is slowed down and spread out that it becomes indistinguishable from the background plasma, which will be warmer because of the energy transferred from the beam [1,2].

The slowing down of a charged particle beam in dusty plasmas can be studied by using some appropriate statistical descriptions, such as the Fokker-Planck collision theory. When a charged particle (as a test particle) flies into the plasma, its velocity distribution $f_T \equiv f_T(\mathbf{v}, t)$ will change due to the collisions with the charged particles of background plasma components. It satisfies the following Fokker-Planck equation [1,2],

$$\frac{\partial f_T}{\partial t} = \sum_j \frac{4\pi q_T^2 q_j^2}{m_T^2} \ln \Lambda \left[ -\frac{\partial}{\partial \mathbf{v}} \cdot \left( f_T \frac{\partial h_j}{\partial \mathbf{v}} \right) + \frac{1}{2} \frac{\partial}{\partial \mathbf{v}} \frac{\partial}{\partial \mathbf{v}} : \left( f_T \frac{\partial^2 g_j}{\partial \mathbf{v} \partial \mathbf{v}} \right) \right], \tag{1}$$

where $q_T$ and $m_T$ are the charge and mass of the test particle, $q_j$ is charge of the *j*th component particle in the background plasma (*j=e, i* and *d* for electrons, ions and dust particles, respectively), and $\ln \Lambda$ is the scattering factor between the test particle and the background plasma. The functions $h_j$ and $g_j$ are expressed respectively by

$$h_j(\mathbf{v}) = \frac{m_T}{\mu_j} \iiint \frac{f_j(\mathbf{v}')}{|\mathbf{v} - \mathbf{v}'|} d\mathbf{v}', \qquad (2)$$

$$g_j(\mathbf{v}) = \iiint f_j(\mathbf{v}') |\mathbf{v} - \mathbf{v}'| d\mathbf{v}', \qquad (3)$$

where $\mu_j = m_T m_j / (m_T + m_j)$ is the reduced mass of the *j*th component particle and $f_j(\mathbf{v}')$ is a stationary velocity distribution of the *j*th component particle in the background plasma. On the right side of Eq.(1), the first term is the friction term, which describes the slowing down of the mean velocity related to $f_T$. While the second term is a diffusion term, which describes the spreading out of the velocity distribution described by $f_T$. Obviously, the two terms are both dependent on the velocity distribution of the particles in the background plasma by the equations (2) and (3). The Fokker-Planck equation describes the particle-particle interactions between the test particle and the charged particles in the background plasma.

The Fokker-Planck collision theory is based on the fact that large-angle deflection of a charged particle in a plasma occurs as a cumulative effect of successive small-angle scatterings rather than as a single binary scattering, because small-angle collisions occur much more frequently than large-angle collisions, and a large-angle deflection of a particle in a plasma is a hundred times more likely to result as a consequence of the many small-angle scatterings that take place as the particle moves through the plasma than as a result of a single binary collision [1,2]. The absorption of a beam particle by the dust particle belongs to the binary collision (the category of large-angle collision), so it does not affect validity of the Fokker-Planck theory.

Traditionally, the velocity distributions of component particles in the background plasma are considered to be a Maxwellian distribution if the plasma is in a thermal equilibrium state [2]. However, Many astrophysical and space plasma observations, such as those on interstellar medium, ionosphere, solar wind, planetary magnetosphere and magnetic field etc. have revealed the universality of non-Maxwellian velocity distributions [3-10], such as the familiar kappa-distribution and the kappa-like distributions in suprathermal space plasmas [3,4], the power-law q-distribution in nonextensive statistics [5], and the non-thermal α-distribution in solitary structures observed by the Freja satellite [6] etc, which in some physical situations have significantly different natures from those plasmas with a Maxwellian distribution. The non-thermal α-distribution is important because it may change the nature of ion sound solitary structures and allow the existence of structures very like those observed [6], and thus widely affect other properties in various types of plasmas including dusty plasmas, such as shear-flow driven tripolar vortices [11], solar wind interaction with dusty plasmas [12], polarized space dusty plasmas [13], dust-acoustic solitary waves [14], shear-flow driven dissipative instability [15], and Landau damping in dust-acoustic waves [16] etc.

If the slowing down of a charged particle beam takes place in the background plasma with a non-Maxwellian velocity distribution, the properties will change depending on the velocity distributions. If the plasma is in a thermal equilibrium state having a Maxwellian velocity distribution, the slowing down can be studied through an error function of average velocity of the particle beam [2]. If the plasma has a velocity *κ*-distribution, the slowing down can be discussed by hypergeometric functions with the κ-parameter [17]. Especially, if the plasma is the nonequilibrium dusty plasma with the velocity κ-distribution, dust particles play a dominant role in the κ-deceleration effects [18]. In this work, we study the slowing down of charged particles in the dusty plasma with the non-thermal α-distribution.

The paper is organized as follows. In section 2, we give the basic slowing down theory and derive the deceleration factor and the slowing down time. In section 3, we make the numerical analysis. In section 4, we give the conclusion.

**2. The slowing down theory**

We consider the nonequilibrium dusty plasma with the non-thermal velocity α-distributions. If we denote $n_e$, $n_i$ and $n_d$ as number density of electrons, ions and dust particles respectively, then they should satisfy the charge neutral condition,

$$\varepsilon_d Z_d n_d + Z_i n_i - n_e = 0, \tag{4}$$

where $Z_d$ and $Z_i$ are charge number of a dust particle and an ion respectively, $\varepsilon_d = +1$ (or $-1$) represents a positively (or negatively) charged dust particle. Generally, we may assume the three plasma components to follow the nonthermal velocity α-distributions respectively, have different temperature $T_j$ and different non-thermal parameter $\alpha_j$, then the velocity distribution functions can be expressed for the $j$th component ($j = e, i, d$) [6,8] by

$$f_{\alpha,j}(\mathbf{v}) = \frac{n_j}{(1+3\alpha_j)(\mathrm{v}_{Tj}\sqrt{2\pi})^3}\left(1+\alpha_j\frac{\mathbf{v}^4}{\mathrm{v}_{Tj}^4}\right)\exp\left(-\frac{1}{2}\frac{\mathbf{v}^2}{\mathrm{v}_{Tj}^2}\right), \tag{5}$$

where $\mathrm{v}_{Tj} = (kT_j/m_j)^{1/2}$ is the thermal speed, $0 \le \alpha_j \le 1$ is a nonthermal parameter which characterizes amounts of high-energy species in the plasma. When $\alpha_j = 0$, the α-distribution function (5) returns to a Maxwellian one.

In the dusty plasma, we consider a beam of test particles with the velocity distribution at time t=0 [1] as

$$f_T(\mathbf{v}, t=0) = \delta(\mathbf{v}-\mathbf{v}_0). \tag{6}$$

The velocity moment of the test particles is defined as the mean velocity, namely,

$$\mathbf{U} = \int \mathbf{v} f_T(\mathbf{v}, t) d\mathbf{v}. \tag{7}$$

According to Ref.[18], by integrating the part of the right-hand side terms of Fokker-Planck equation (1), we can obtain the velocity moment equation (i.e. the deceleration factor, $\partial \mathbf{U}/\partial t$). And then substituting the function (6) into the deceleration factor, $\partial \mathbf{U}/\partial t$, we can study the slowing down of the particle beam at time t=0 in the dusty plasma by the following moment equation,

$$\left.\frac{\partial \mathbf{U}}{\partial t}\right|_{t=0} = \frac{4\pi q_T^2 \ln \Lambda}{m_T^2} \sum_j q_j^2 \frac{\partial}{\partial \mathbf{U}} h_{\alpha,j}(\mathbf{U}), \tag{8}$$

where $\mathbf{U}$ is the average velocity $\mathbf{v}_0$ at t=0, and the function is now expressed by the non-thermal α-distribution as

$$h_{\alpha,j}(\mathbf{v}) = \frac{m_T}{\mu_j}\iiint \frac{f_{\alpha,j}(\mathbf{v}')}{|\mathbf{v}-\mathbf{v}'|}d\mathbf{v}' = \frac{m_T n_j}{\mu_j(2\pi)^{3/2}(1+3\alpha_j)\mathrm{v}_{Tj}^3}\iiint \frac{d\mathbf{v}'}{|\mathbf{v}-\mathbf{v}'|}\left(1+\alpha_j\frac{\mathbf{v}'^4}{\mathrm{v}_{Tj}^4}\right)\exp\left(-\frac{1}{2}\frac{\mathbf{v}'^2}{\mathrm{v}_{Tj}^2}\right). \tag{9}$$

In order to calculate (9), we can let $\xi = |\mathbf{v} - \mathbf{v}'| = \sqrt{v^2 + v'^2 - 2vv'\cos\theta}$, where $\theta$ is the angle between $\mathbf{v}$ and $\mathbf{v}'$, then we have $\cos\theta = \dfrac{v^2 + v'^2 - \xi^2}{2vv'}$, $\sin\theta d\theta = -d\cos\theta = \dfrac{\xi d\xi}{vv'}$ and $d\mathbf{v}' = v'^2 dv' \sin\theta d\theta d\varphi = \dfrac{v'}{v} dv' \xi d\xi d\varphi$. The integral range of $\xi$ is from $|v - v'|$ to $v + v'$, so Eq.(9) becomes

$$h_{\alpha,j}(\mathbf{v}) = \frac{2\pi}{v}\frac{m_T}{\mu_j}\int_0^\infty v' f_{\alpha,j}(\mathbf{v}') dv' \int_{|v-v'|}^{v+v'} d\xi = \frac{2\pi}{v}\frac{m_T}{\mu_j}\int_0^\infty v'(v + v' - |v - v'|) f_{\alpha,j}(\mathbf{v}') dv'$$

$$= \frac{m_T n_j}{v(2\pi)^{1/2} \mu_j (1 + 3\alpha_j) v_{Tj}^3} \int_0^\infty v'(v + v' - |v - v'|) \left(1 + \alpha_j \frac{v'^4}{v_{Tj}^4}\right) \exp\left(-\frac{1}{2}\frac{v'^2}{v_{Tj}^2}\right) dv', \quad (10)$$

where $v + v' - |v - v'| = 2v$ for $v' > v$ and $2v'$ for $v' < v$. Thus we can get that

$$h_{\alpha,j}(\mathbf{v}) = \frac{\sqrt{2} n_j m_T}{(1 + 3\alpha_j)\mu_j \sqrt{\pi} v_{Tj}^3}\left[\int_v^\infty v'\left(1 + \alpha_j \frac{v'^4}{v_{Tj}^4}\right)\exp\left(-\frac{1}{2}\frac{v'^2}{v_{Tj}^2}\right)dv' + \frac{1}{v}\int_0^v v'^2\left(1 + \alpha_j \frac{v'^4}{v_{Tj}^4}\right)\exp\left(-\frac{1}{2}\frac{v'^2}{v_{Tj}^2}\right)dv'\right]$$

$$= \frac{-m_T n_j}{\mu_j \sqrt{\pi}(1 + 3\alpha_j)}\left[\alpha_j\left(7 + v^2 \frac{m_j}{kT_j}\right)\sqrt{\frac{2m_j}{kT_j}}\exp\left(-\frac{m_j v^2}{2kT_j}\right) - \frac{\sqrt{\pi}}{v}(1 + 15\alpha_j) erf\left(v\sqrt{\frac{m_j}{2kT_j}}\right)\right], \quad (11)$$

where $erf(\ldots)$ is an error function.

Now, substituting Eq.(11) into Eq.(8), we can obtain the velocity moment (i.e., the deceleration factor) of the test particle in the dusty plasma with the non-thermal α-distribution by

$$\left.\frac{\partial \mathbf{U}}{\partial t}\right|_{t=0} = \frac{4\sqrt{\pi} q_T^2 \ln\Lambda}{m_T^2 U^2}\sum_j \frac{n_j q_j^2}{(1 + 3\alpha_j)}\left(1 + \frac{m_T}{m_j}\right)\left\{\left[\alpha_j U^4\left(\frac{m_j}{kT_j}\right)^2 + 5\alpha_j U^2\left(\frac{m_j}{kT_j}\right) + 15\alpha_j + 1\right] U\sqrt{\frac{2m_j}{kT_j}}\exp\left(-\frac{m_j U^2}{2kT_j}\right)\right.$$

$$\left. - \sqrt{\pi}(1 + 15\alpha_j) erf\left(U\sqrt{\frac{m_j}{2kT_j}}\right)\right\}. \quad (12)$$

The test particle may be an electron, an ion, or a dust grain. The slowing down time of a charged particle beam in the dusty plasma is defined [2] by

$$\tau_{\alpha,s} = -U\left(\left.\frac{\partial \mathbf{U}}{\partial t}\right|_{t=0}\right)^{-1}. \quad (13)$$

And therefore, we have that

$$\tau_{\alpha,s} = \frac{-U^3 m_T^2}{4\sqrt{\pi} q_T^2 \ln\Lambda}\left\{\sum_j \frac{n_j q_j^2}{(1 + 3\alpha_j)}\left(1 + \frac{m_T}{m_j}\right)\left[\left(\alpha_j U^4\left(\frac{m_j}{kT_j}\right)^2 + 5\alpha_j U^2\left(\frac{m_j}{kT_j}\right) + 15\alpha_j + 1\right) U\sqrt{\frac{2m_j}{kT_j}}\exp\left(-\frac{m_j U^2}{2kT_j}\right)\right.\right.$$

$$\left.\left. - \sqrt{\pi}(1 + 15\alpha_j) erf\left(U\sqrt{\frac{m_j}{2kT_j}}\right)\right]\right\}^{-1}. \quad (14)$$

When we take $\alpha_j = 0$, it returns to that in the plasma with a Maxwellian distribution [19],

$$\tau_{0,s} = \frac{U^3 m_T^2}{4\pi q_T^2 \ln \Lambda} \left\{ \sum_j n_j q_j^2 \left(1+\frac{m_T}{m_j}\right) \left[ -U\sqrt{\frac{2m_j}{\pi k T_j}} \exp\left(-\frac{m_j U^2}{2kT_j}\right) + erf\left(U\sqrt{\frac{m_j}{2kT_j}}\right) \right] \right\}^{-1}. \quad (15)$$

### 3. Numerical analyses

In order to study the deceleration characteristics further of a charged particle beam in the dusty plasma with the non-thermal velocity α-distribution, now we make the numerical analysis. The deceleration factor about each plasma component will be analyzed. We will focus on the role of the α-parameter in the deceleration factor Eq.(12), which can be written into the following form about the electron term, the ion term and the dust term respectively,

$$\left.\frac{\partial \mathbf{U}}{\partial t}\right|_{t=0} = \sum_j \left(\frac{\partial \mathbf{U}}{\partial t}\right)_j = \left(\frac{\partial \mathbf{U}}{\partial t}\right)_e + \left(\frac{\partial \mathbf{U}}{\partial t}\right)_i + \left(\frac{\partial \mathbf{U}}{\partial t}\right)_d, \quad (16)$$

where

$$\left(\frac{\partial \mathbf{U}}{\partial t}\right)_j = \frac{4\sqrt{\pi} q_T^2 \ln \Lambda}{m_T^2 U^2} \frac{n_j q_j^2}{(1+3\alpha_j)} \left(1+\frac{m_T}{m_j}\right) \left\{ \left[\alpha_j U^4 \left(\frac{m_j}{kT_j}\right)^2 + 5\alpha_j U^2 \left(\frac{m_j}{kT_j}\right) + 15\alpha_j + 1\right] U\sqrt{\frac{2m_j}{kT_j}} \exp\left(-\frac{m_j U^2}{2kT_j}\right) \right.$$

$$\left. -\sqrt{\pi}(1+15\alpha_j) erf\left(U\sqrt{\frac{m_j}{2kT_j}}\right) \right\}. \quad (17)$$

Eqs.(17) for $j=e, i, d$ are the deceleration factors due to the electrons, the ions and the dust grains, respectively, in the dusty plasma with the non-thermal velocity α-distribution. We can see that the slowing down effects depend on average velocity, charge, mass of the beam particles, but also depends on the physical quantities such as temperatures, masses, densities, and charges of the dusty plasma components. Especially, the effects are significantly dependent on the non-thermal parameters $\alpha_j$ of the dusty plasma components.

Although these physical quantities in astrophysical and space plasmas are quite uncertain, the typical values are usually found in some plasma space environments [20-25]. For example, the data of dusty plasma in Saturn's E-ring can be used for the calculations of the slowing down effects [18]. The scattering factor lnΛ is set as a constant between 8~25, because it does not affect the effect of the non-thermal α-distribution on the property of $\partial \mathbf{U}/\partial t|_{t=0}$. As an example, in the following Figures, we will show that when an electron beam passes through the non-thermal dusty plasma, the deceleration factors Eqs.(17) for $j=e, i, d$ are respectively as a function of U for several different values of the α-parameter. In addition, we also plot the slowing down time τ as a function of U on the basis of Eq.(14).

In Fig. 1, based on Eq.(17), we show the contribution of the electron components to the deceleration factor for five different values of the non-thermal parameter $\alpha_e$. In Fig. 2, based on Eq.(17), we show the contribution of the ion components to the deceleration factor for five different values of the non-thermal parameter $\alpha_i$. In Fig. 3, based on Eq.(17), we show the contribution of the dust components to the deceleration factor for five different values of the non-thermal parameter $\alpha_d$. In Fig. 4, based on Eq.(14), we show the slowing down time as a function of the mean velocity of the electron beam for five different values of the non-thermal α-parameter in the dusty plasma. In the all calculations above, the non-thermal parameter $\alpha_j=0$ corresponds to the cases with a Maxwellian velocity distribution.

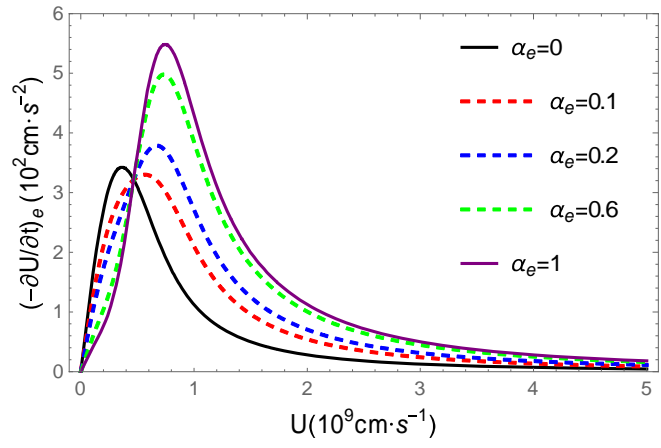

Fig.1

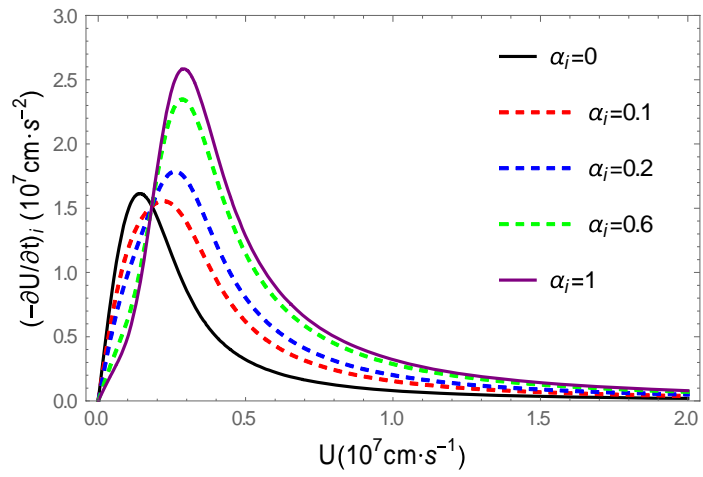

Fig.2

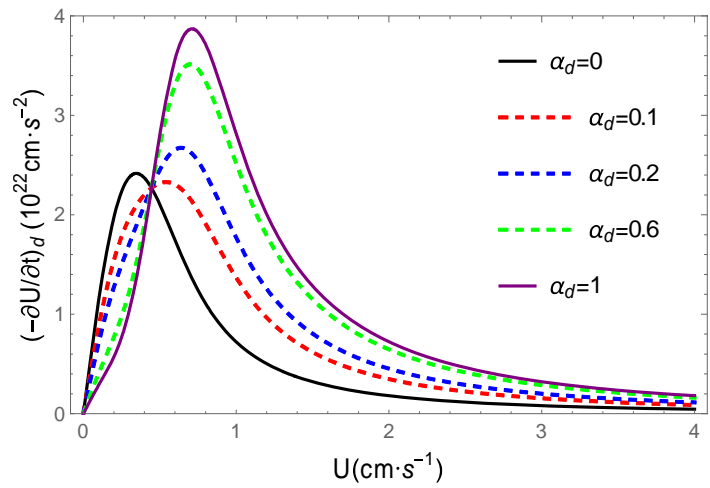

Fig.3

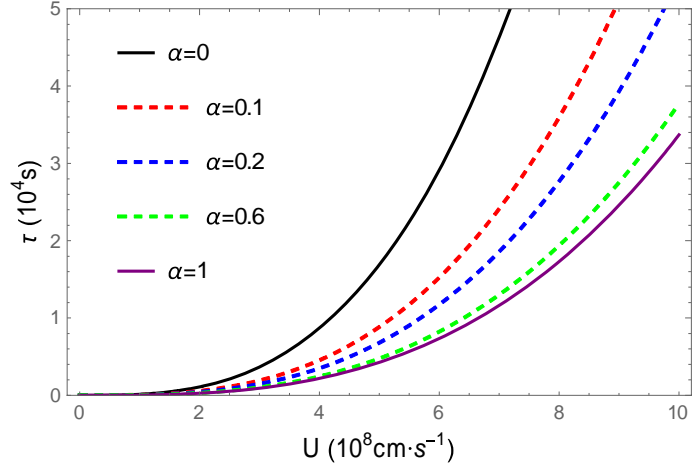

Fig.4

From Fig.1, Fig. 2 and Fig.3, we find the same characteristic that the non-thermal α-distributions of the plasma components have a significant effect on the deceleration factors. With increase of the mean velocity U, the deceleration factor increases rapidly, reaches a peak and then decreases gradually. And the peak of the deceleration factor moves generally to the right with the increase of the α-parameter.

From Fig.4, we find that the slowing down time will decrease with the increase of the non-thermal α-parameter, and so the slowing down time of a charged particle beam in a non-thermal dusty plasma is generally less than that in a Maxwellian one.

## 4. Conclusion

In conclusion, we have used the Fokker-Planck collision theory to study the slowing down of charged particle beam in the dusty plasmas with a non-thermal velocity α-distribution. The three plasma components, electrons, ions and dust particles, can have different non-thermal α-parameters. We derived the slowing down deceleration factor and the slowing down time, given by Eq.(12) and Eq.(14) respectively. When we take the non-thermal α=0, the above formula recovers to those with a Maxwellian velocity distribution.

We have taken the data of Saturn's E-ring [18] dusty plasma as an example to make numerical analysis of the deceleration factor and the slowing down time. We showed that the non-thermal α-distributions of the plasma components have a significant effect on the deceleration factors and slowing down time. With increase of the mean velocity U, the deceleration factor increases rapidly, reaches a peak and then decreases gradually. And the entire peak of the deceleration factor moves generally to the right with the increase of the α-parameter. The slowing down time decreases with the increase of the non-thermal α-parameter, and so the slowing down time of a charged particle beam in the non-thermal dusty plasma is generally less than that in a Maxwellian one.

**Acknowledgements**

This work was supported by the National Natural science foundation of China under Grant No. 11775156.